\documentclass[conference]{IEEEtran}
\IEEEoverridecommandlockouts

\usepackage{cite}
\usepackage{amsmath,amssymb,amsfonts}
\usepackage{algorithmic}
\usepackage{graphicx}
\usepackage{textcomp}
\usepackage{xcolor}

\usepackage{multirow}
\usepackage{float}
\usepackage{array}
\usepackage{subfig}
\usepackage{amssymb}

\def\BibTeX{{\rm B\kern-.05em{\sc i\kern-.025em b}\kern-.08em
    T\kern-.1667em\lower.7ex\hbox{E}\kern-.125emX}}
\begin{document}

\title{Multi-modal Learning with Missing Modality in Predicting Axillary Lymph Node Metastasis\\
\thanks{This study was partially supported by the National Natural Science Foundation of China (Grant no. 92270108), Zhejiang Province Natural Science Foundation of China (Grant no. XHD23F0201).}
}

\author{\IEEEauthorblockN{Shichuan Zhang$^{\star \dagger}$ \qquad Sunyi Zheng$^{\dagger}$ \qquad Zhongyi Shui$^{\star \dagger}$ \qquad Honglin Li$^{\star \dagger}$ \qquad Lin Yang$^{\dagger}$}
\IEEEauthorblockA{\textit{$^{\star}$ Zhejiang University, Hangzhou, Zhejiang, China} \\
\textit{$^{\dagger}$ School of Engineering, Westlake University, Hangzhou, Zhejiang, China}\\
zhangshichuan@westlake.edu.cn, zhengsunyi@westlake.edu.cn, yanglin@westlake.edu.cn}
}


\maketitle

\begin{abstract}
Multi-modal Learning has attracted widespread attention in medical image analysis. Using multi-modal data, whole slide images (WSIs) and clinical information, can improve the performance of deep learning models in the diagnosis of axillary lymph node metastasis. However, clinical information is not easy to collect in clinical practice due to privacy concerns, limited resources, lack of interoperability, etc.
Although patient selection can ensure the training set to have multi-modal data for model development, missing modality of clinical information can appear during test. This normally leads to performance degradation, which limits the use of multi-modal models in the clinic. To alleviate this problem, we propose a bidirectional distillation framework consisting of a multi-modal branch and a single-modal branch. The single-modal branch acquires the complete multi-modal knowledge from the multi-modal branch, while the multi-modal learns the robust features of WSI from the single-modal. We conduct experiments on a public dataset of  Lymph Node Metastasis in Early Breast Cancer to validate the method. Our approach not only achieves state-of-the-art performance with an AUC of 0.861 on the test set without missing data, but also yields an AUC of 0.842 when the rate of missing modality is 80\%. This shows the effectiveness of the approach in dealing with multi-modal data and missing modality. Such a model has the potential to improve treatment decision-making for early breast cancer patients who have axillary lymph node metastatic status. 
\end{abstract}

\begin{IEEEkeywords}
Missing modality, Whole slide image,  Clinical data.
\end{IEEEkeywords}

\section{Introduction}
Breast cancer has become the most deadly disease for women worldwide. The prediction of axillary lymph node metastasis(ALNM) can guide treatment, therefore is crucial to improve the survival rate of early breast cancer patients. Previous works \cite{hu2021deep,zhao2020predicting,harmon2020multiresolution} have been devoted to the prediction of LNM. Li, et al \cite{li2021multi} combine the histopathological images and tabular clinical data, including age, gender and tumor location, to improve the performance of ALNM prediction. Besides, multi-modal learning \cite{dalmaz2022resvit,zhang2023multi,acosta2022multimodal} in other medical fields have also achieved remarkable results. A multi-modal Transformer \cite{zheng2022multi} is introduced for the survival prediction of nasopharyngeal carcinoma patients. Hong, et al \cite{hong2021predicting} combine the clinical features and histological images to predict the molecular subtypes and mutation status. Generally, existing research efforts mainly focus on how to fuse multi-modal data effectively.


\begin{figure}[t]
\centering
	\subfloat[]{\includegraphics[width = 0.24\textwidth]{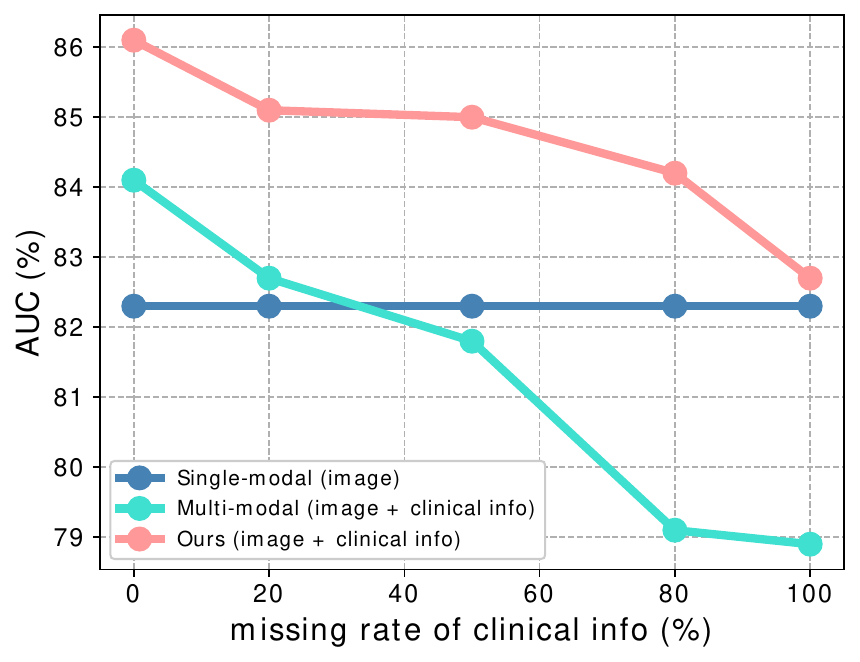}}
	\subfloat[]{\includegraphics[width = 0.24\textwidth]{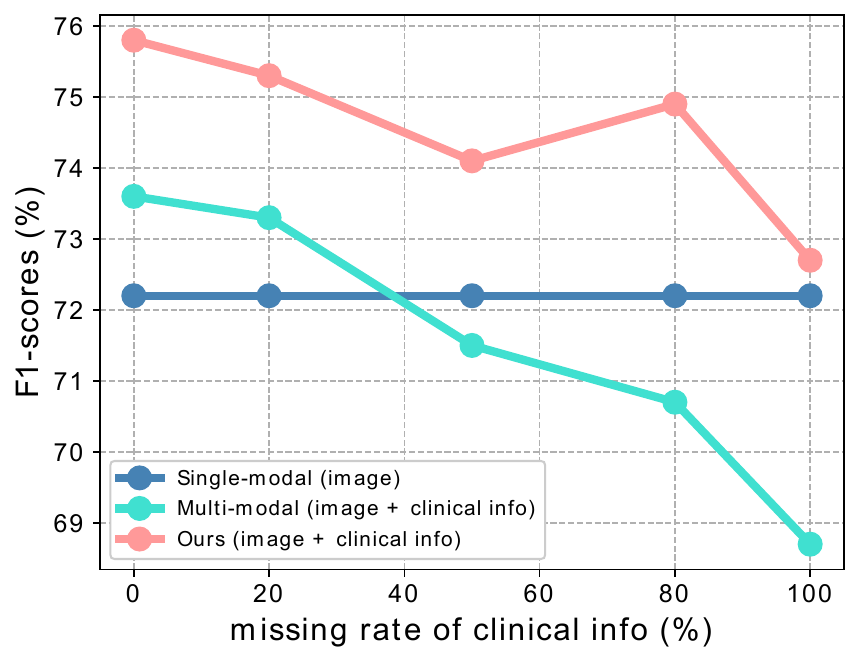}}
\caption{Model performance at different missing rates of clinical information. The multi-modal model \cite{xu2021predicting} has better results than that of the single-modal model trained on images at a low missing rate of clinical data, but the situation is reversed when the missing rate is high. By using bidirectional distillation, our multi-modal method with the same backbone can achieve good performance regardless of missing rates.}
\label{pre-experiment}
\end{figure}

\begin{figure*}[t]
\centering
\includegraphics[width=0.88\textwidth]{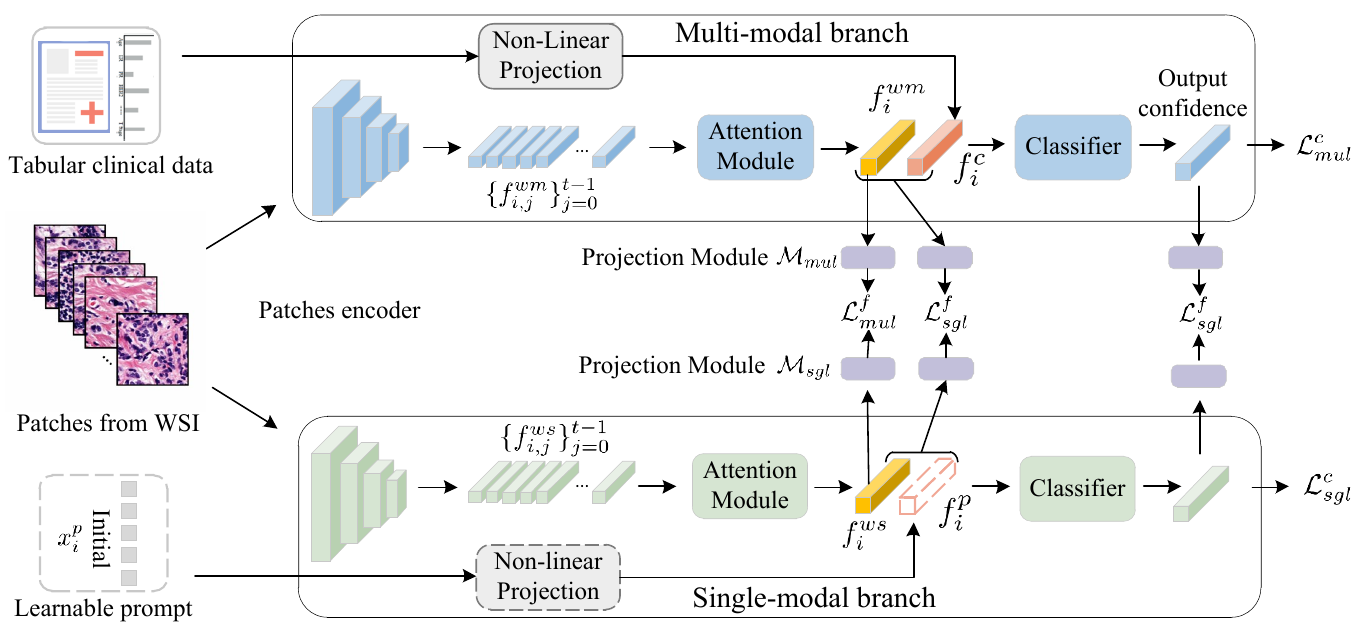}
\caption{An overview of the bidirectional distillation framework.} \label{overview}
\end{figure*}

Studies have been designed to tackle the missing modality problem. Transfer learning \cite{zheng2021deep,garcia2018modality,xing2022discrepancy,zhang2021modality} is effective in handling the fully missing modality problem. However previous methods ignore a practical situation that one of the modalities is often partly missing. It poses a different challenge with the fully missing modality \cite{rahate2022multimodal}. Ma, et al \cite{ma2021smil} consider various missing situations and leverage a generative model to produce missing text during training and testing. Nevertheless, generative models require a large number of training pairs. 
The combined application of histology and molecular markers is employed for the classification of diffuse glioma through multi-task learning \cite{wang2023multi}. However, the gaps in method performance attributed to distinct modalities cannot be avoided.
To our best knowledge, there are no efforts on multi-modal learning with both partly and fully missing modality for pathology images and clinical data. 

Making full use of comprehensive information such as histopathological images and clinical data can effectively improve the performance of deep learning models \cite{hohn2021combining,yang2022prediction,huang2022predicting}. But clinical data are not always available in the real world due to privacy concerns or limited resources, especially in actual testing. Therefore, the question about will the fusion of multi-modal information in training helps even if the task is single-modal or partly missing modality at test time remains. As shown in Fig.\ref{pre-experiment}, modality missing at the testing phase can seriously affect model performance. The performance of the model \cite{xu2021predicting} that learns from multi-modal data even becomes worse than that of the model learning from single-modal data when considering severe missing of clinical data. Thus, the problem needs to be solved: how to learn a multi-modal model from a complete training dataset while it is robust to fully or partly missing modality during testing.

In order to take full advantage of the clinical data in training set effectively and face various missing styles (partly missing and fully missing) during testing flexibly, we propose a bidirectional distillation (BD) framework as shown in Fig.\ref{overview}. Our contributions can be concluded as follows:

\begin{itemize}
    \item We propose a BD framework consisting of a single-modal branch and a multi-modal branch, which can flexibly tackle modality complete or incomplete inputs in a unified manner by turning off or on the single branch when testing.
    \item In order to transfer the knowledge of clinical information to the single-modal branch, we introduce a learnable prompt during the distillation from the multi-modal branch to the single-modal branch.
    \item The learning of complicated fused features may lead to the overfitting on the feature learning of WSI \cite{zadeh2017tensor}, which is verified in the experiment. To tackle this challenge, we leverage the distillation from the single-modal branch to the multi-modal branch to extract robust features of WSI in the multi-modal branch.
    \item We additionally conduct further research on the missing modality within WSIs. The experimental results demonstrate the strong performance of our method regardless of the missing modality.
\end{itemize}

\section{Methodology}
\subsection{Problem Formulation.}
Missing modality of clinical data in the test time is considered in the paper. The dataset is divided into a training set and test set: $\mathcal{D} = \{ \mathcal{D}^d, \mathcal{D}^v\}$. We consider the training set $\mathcal{D}^d = \{(x^w_i, x^c_i, y_i)\}_{i=0}^{n-1}$ as a modality-complete dataset, where $x^w_i$ and $x^c_i$ represent two different modalities (whole slide images (WSI) and clinical information) of the $i$-th sample, $y_i$ is the corresponding label and $n$ is the total number of the samples in the training set. The test set $\mathcal{D}^v = \{(x^w_0, x^c_0, y_0),(x^w_1, y_1),...\}$ is a modality-incomplete dataset. There exist samples in the data set $\mathcal{D}^v$ that do not contain the clinical information. In this paper,
we aim to make full use of the multi-modal information in training set to improve the model performance and flexibly deal with the modality-missing problem in the test set.

\subsection{Multi-modal Branch Learning.}
Specifically, a WSI $x^w_i$ is divided into $t$ small patches which feed into an encoder. In the multi-modal branch, deep features $\{f_{i,j}^{wm}\}_{j=0}^{t-1}$ from the encoder are aggregated to the fused feature $f_i^{wm}$ by simple attention \cite{vaswani2017attention}. 
\begin{equation}
f_i^{wm} = \sum_j \{f_{i,j}^{wm}\}_{j=0}^{t-1} \cdot \mathcal{H}(\{f_{i,j}^{wm}\}_{j=0}^{t-1}),
\end{equation}
where $\mathcal{H}(\cdot)$ is a non-linear projection function whose parameters are learnable. The output of $\mathcal{H}(\cdot)$ is a 1D vector with length $t$. The $j$th element corresponds to the patch feature $f_{i,j}^{wm}$ during summation. We combine $f_i^{wm}$ and the mapped feature $f_i^c$ of clinical data to calculate the final classification loss $\mathcal{L}_{mul}^c$. 
\begin{equation}
\mathcal{L}_{mul}^c = -\sum_i y_i log(\mathcal{G}_{mul}([f_i^{wm}, f_i^c])),
\end{equation}
where $\mathcal{G}_{mul}(\cdot)$ is the final classifier in the multi-modal branch and $[\cdot, \cdot]$ is a feature splicing operation.

In order to avoid the feature learning of WSI being affected by clinical information, we transfer the knowledge from the single-modal branch to the multi-modal branch. We define the intermediate output $\{f_k^{mul}\}_{k=0}^{l-1}$ of the deep layers in the multi-modal branch and $\{f_k^{sgl}\}_{k=0}^{l-1}$ in the single-modal branch. $l$ is the number of deep feature layers in the network. In this paper, we choose the WSI features from the final layer $f_{l-1}^{mul}$ and $f_{l-1}^{sgl}$ (the output of the attention module $f_i^{wm}$ and $f_i^{ws}$ in Fig. \ref{overview}), which exhibit the most robust semantics of WSIs. 
The knowledge distillation for the WSI features can be represented as follows:
\begin{equation}
    \mathcal{L}_{mul}^f = \sum_i \mathcal{D}(\mathcal{M}_{mul}(f_{i}^{wm}), \mathcal{M}_{sgl}(f_{i}^{ws})),
\end{equation}
where $\mathcal{D}$ is a distance function that measures the gap of features between the single-modal and multi-modal branches. We choose the mean square error as the distance measure function. $\mathcal{M}_{mul}$ and $\mathcal{M}_{sgl}$ are the projection modules that can transfer the intermediate output feature to the target representation. 
The total loss function for the learning of the multi-modal branch is
\begin{equation}
    \mathcal{L}_{mul} = \mathcal{L}_{mul}^c + \lambda_m * \mathcal{L}_{mul}^f,
\end{equation}
where $\lambda_m$ is a hyper-parameter to weigh different items. We utilize the classification loss $\mathcal{L}_{mul}^c$ and the distillation loss $\mathcal{L}_{mul}^f$ to update the multi-modal branch simultaneously.

\subsection{Single-modal Branch Learning.}
Following patch fusion steps described in Multi-modal Branch Learning, we convert deep features $\{f_{i,j}^{ws}\}_{j=0}^{t-1}$ to $f_i^{ws}$.
We employ a learnable prompt \cite{chen2022knowprompt} $x_i^p$ to signal the single-model branch when missing modality and memorize the missing information of clinical data $x_i^c$. We map $x_i^p$ to a feature $f_i^p$ by a non-linear function. The dimension of $f_i^p$ is as same as that of the feature $f_i^c$ in the multi-modal branch. We then combine the WSI feature $f_i^{ws}$ and the prompt feature $f_i^p$. Afterward, the knowledge of clinical data is transferred from the multi-modal branch to the single-modal branch based on the distillation loss:
\begin{equation}
  \begin{aligned}
      \mathcal{L}_{sgl}^f = &\mathcal{D}([f_i^{ws}, f_i^p], [f_i^{wm}, f_i^c]) + \\
    &KL(\mathcal{G}_{sgl}([f_i^{ws}, f_i^p]), \mathcal{G}_{mul}([f_i^{wm}, f_i^c]).
  \end{aligned}
\end{equation}
We also apply mean square error for $\mathcal{D}$. $KL(\cdot, \cdot)$ \cite{beyer2022knowledge} is the KL divergence function for the predicted confidence. $\mathcal{G}_{sgl}$ and $\mathcal{G}_{mul}$ are two final classifiers for single-modal branch and multi-modal branch, respectively. The loss function $\mathcal{L}_{sgl}^f$ is only used for the learning of the prompt as shown in Fig.\ref{overview}. 

Similarly, there is also a classification loss $\mathcal{L}_{sgl}^c$ for the learning of WSIs in the single-modal branch. Consequently, the total loss function for the training of a single-modal branch is presented as follows:
\begin{equation}
    \mathcal{L}_{sgl} = \mathcal{L}_{sgl}^c + \lambda_s * \mathcal{L}_{sgl}^f
\end{equation}
The loss function is used for the update of the single-modal branch while the multi-modal branch is frozen. 
During testing, the BD framework can tackle modality complete or incomplete inputs in a unified manner by turning off or on the single branch.


\section{Experiments and Results} 

\begin{figure*}[t]
\centering
	\subfloat[Filling]{\includegraphics[width = 0.29\textwidth]{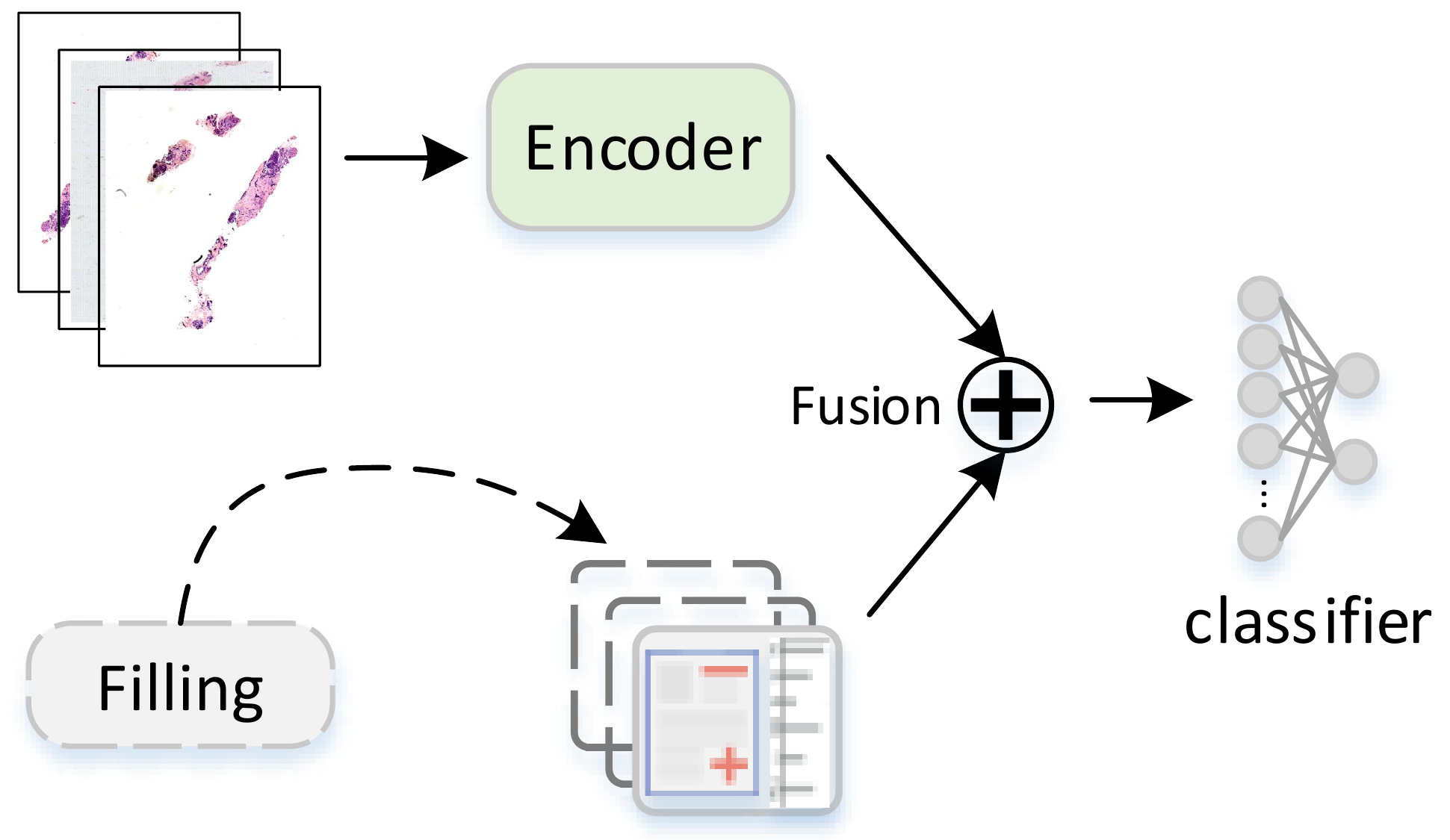}}
	\hspace{4mm}
	\subfloat[AE]{\includegraphics[width = 0.29\textwidth]{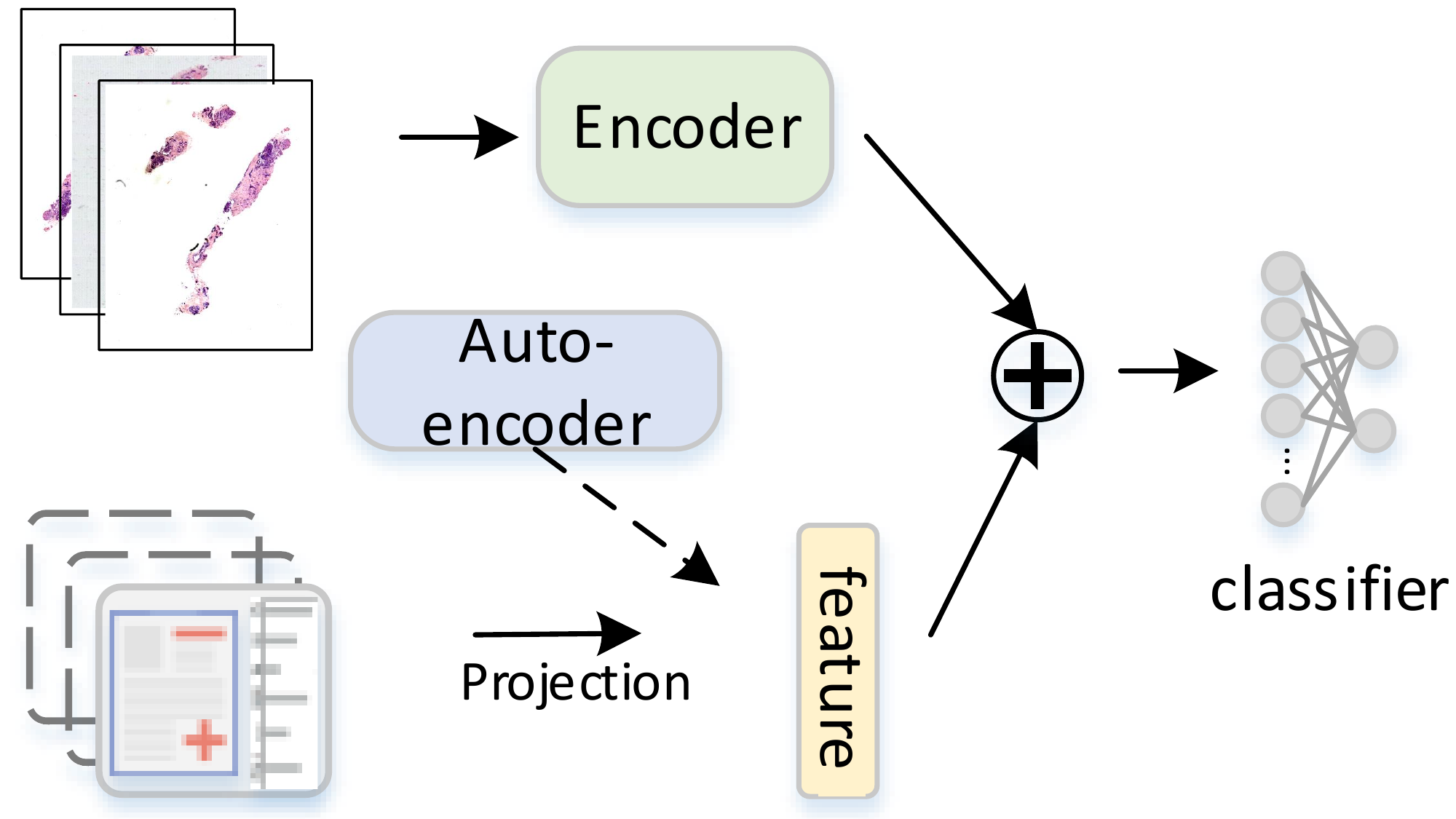}}
	\hspace{4mm}
	\subfloat[Ensemble]{\includegraphics[width = 0.29\textwidth]{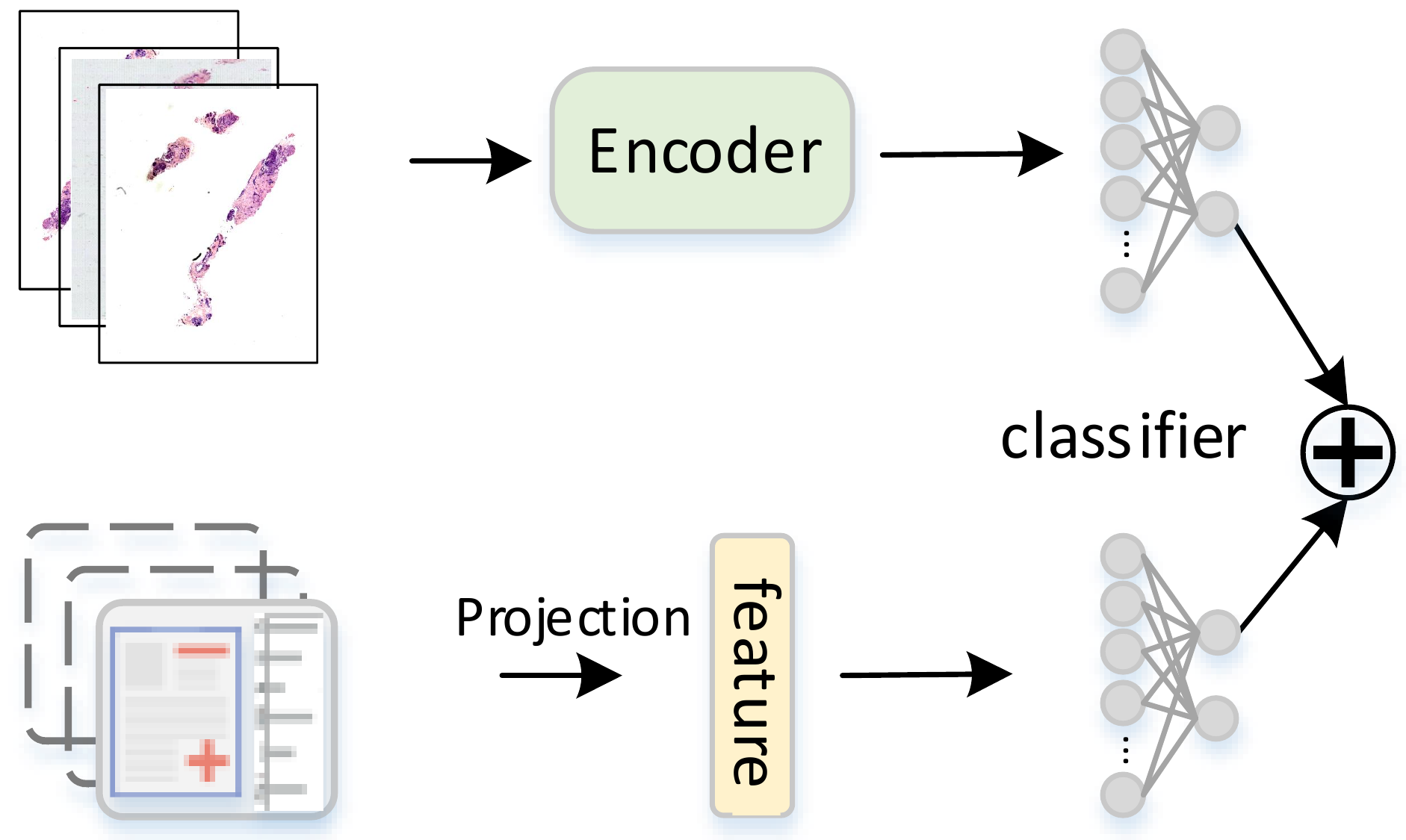}} 
\caption{Structures of the methods for multi-modal learning with missing modality}
\label{baselines}
\end{figure*}

\subsection{Dataset and Experimental Settings.}
The experimental dataset is from a grand challenge named Early Breast Cancer Core-Needle Biopsy WSI (BCNB) \cite{xu2021predicting}. Paired multi-modal data containing WSIs and clinical information is provided by the dataset. All WSIs are hematoxylin and eosin stained and the clinical data consists the information of age, tumor size, ER, PR and HER2. We use the information to predict the metastatic status ($N_0$ and $N_+$) of axillary lymph nodes. Since it is a binary classification task, we use the metrics Area Under Curve (AUC) and F1-scores (F1) to validate the proposed method. F1 represents the averaged results in the prediction of metastatic status.

We randomly split the dataset into a training and test set with 80\% and 20\%. A subset with 20\% is separated from the training set for validation. We assume that the training set is complete with paired modalities. While the clinical data in the test set can be missed at a random rate. In the training process of our method, stochastic gradient descent with a momentum of 0.3, a weight decay rate of $1 \times 10^{-3}$ serves as the optimizer. The learning rate is initialized at $1 \times 10^{-4}$. The hyper-parameters $\tau$, $\lambda_s$ and $\lambda_m$ are set to 1.2, 0.5 and 0.6, respectively. We initialize the learnable prompt with a length of 50. Early stopping is used to avoid overfitting by monitoring the F1 scores in the training set. The code is implemented based on python3 and pytorch-1.9 and all experiments are conducted using NVIDIA A100 GPUs.

All non-linear and linear projection modules are composed of fully connected layers and the ReLU non-linear activation function. The function $\mathcal{H}(\cdot)$ within the attention module consists of two hidden layers with corresponding activation functions. We employ two layers with hidden sizes of 100 and 50 to map the learnable prompt $x_i^p$ to $f_i^p$. Both $\mathcal{M}_{mul}$ and $\mathcal{M}_{sgl}$ are fully connected layers used to map features to a dimension of 64.

\begin{table}[t]
\caption{Results of our method (BD) and other methods. We regard the method Filling as a baseline and calculate the changes ($\Delta$) at various missing ratios.}
\begin{center}
\begin{tabular}{|c|c|c|c|c|c|}
\hline
Missing rate(\%)& Methods& AUC& $\Delta$ AUC& F1& $\Delta$ F1\\
\hline
\multirow{2}{*}{0} & image only & 82.3& - & 72.2& - \\
 & clinical only &71.6 & - &62.2 & - \\
\hline
\multirow{5}{*}{0} & Filling &84.1 & &73.6 & \\ 
& AE &84.1 &0.0 &73.6 &0.0\\
& Ensemble &\underline{85.1} &1.0 &\underline{74.0} &0.4\\
& SMIL &82.6 &-1.5 &72.7 &-0.9 \\
& BD &\textbf{86.1} &2.0 &\textbf{75.8} &2.2 \\
\hline
\multirow{5}{*}{50} & Filling &81.8 & &71.5 & \\
& AE &81.3 &-0.5 &71.5 &0.0 \\
& Ensemble &\underline{83.7} &1.9 &73.0 &1.5 \\
& SMIL &82.2 &0.4 &\underline{73.2} &1.7 \\
& BD &\textbf{85.0} &3.2 &\textbf{74.1} &2.6 \\
\hline
\multirow{5}{*}{80}& Filling &79.1 & &70.7 & \\
& AE &79.9 &0.8 &70.6 &-0.1 \\
& Ensemble &\underline{82.8} &3.7 &\underline{71.7} &1.0 \\
& SMIL &80.0 &0.9 &71.5 &0.8 \\
& BD &\textbf{84.2} &5.1 &\textbf{74.9}&4.2 \\
\hline
\multirow{5}{*}{100}& Filling &78.9 & &68.7 & \\
& AE &79.6 &0.7 &69.4 &0.7 \\
& Ensemble &\underline{82.3} &3.4 &\underline{72.2} &3.5 \\
& SMIL &78.8 &-0.1 &69.7 &1.0 \\
& BD &\textbf{82.7} &3.8 &\textbf{72.7} &4.0 \\
\hline
\end{tabular}
\label{comparison}
\end{center}
\end{table}

\begin{table}[t]
\caption{Ablation study on the effect of the two parts $\mathcal{S}\rightarrow \mathcal{M}$ (learning from single-modal branch) and $\mathcal{M}\rightarrow \mathcal{S}$ (learning from multi-modal branch).}
\begin{center}
\begin{tabular}{|c|c|c|c|}
\hline
\makebox[0.20\linewidth][c]{Missing\ rate(\%)} & \makebox[0.14\linewidth][c]{$\mathcal{S}\rightarrow \mathcal{M}$}& 
\makebox[0.14\linewidth][c]{$\mathcal{M}\rightarrow \mathcal{S}$}& 
\makebox[0.12\linewidth][c]{F1-score}\\
\hline
 \multirow{3}{*}{0}&  & &74.2 \\
 \cline{2-4} 
 & \checkmark & &75.8 \\
 \cline{2-4} 
 &  &\checkmark &74.2 \\
 \hline
 \multirow{3}{*}{80}& & &72.0 \\
 \cline{2-4} 
 & \checkmark & &72.1 \\
 \cline{2-4} 
 &  &\checkmark &73.8 \\
\hline
\end{tabular}
\label{ablation}
\end{center}
\end{table}

\subsection{Comparison with other methods.}
We compared our proposed approach with representative methods (AE \cite{dumpala2019audio}, Ensemble \cite{zhang2019hierarchical}, Filling, SMIL \cite{ma2021smil}) in dealing with the missing modality problem for multi-modal learning, The mechanism of first three intuitive methods are as shown in Fig. \ref{baselines}.

\begin{itemize}
    \item \textit{Filling} is the method that aims to fill the missing clinical data with zero vectors. The model structure is based on the model LNMP \cite{xu2021predicting}. It is the same as LNMP when the modalities are complete during test.
    \item \textit{AE} is designed to generate the missed deep features of clinical data automatically. This model is trained with two stages. First, we train an LNMP model with the modality-complete training set. Then, an auto-encoder is trained to generate missed features, the input and output of which are features of the WSIs and clinical data, respectively. 
    \item \textit{Ensemble} is the model that has two individual networks. One is the WSI recognition network, whose output is the predicted probability. The other one is the classification network for clinical data. We get the final prediction result by fusing the probabilities from the two networks. We only use the first network if there is no input of clinical data.
\end{itemize}

\begin{table}[t]
\caption{Ablation study on $\lambda_m$ and $\lambda_s$. The last column means the epoch of saved best model during training.}
\begin{center}
\begin{tabular}{|c|c|c|c|c|}
\hline
Missing rate(\%)& \textbf{$\lambda_m$}& \textbf{$\lambda_s$}& F1-score & Epoch\\
\hline
\multirow{8}{*}{100}& 0.2& 0.5& 71.7& 30\\
\cline{2-5} 
 & 0.4& 0.5& 70.3& 26\\
\cline{2-5} 
 & 0.6& 0.5& 72.7& 31\\
\cline{2-5} 
 & 0.8& 0.5& 72.2& 28\\
\cline{2-5} 
 & 0.6& 0.2& 71.0& 33\\
\cline{2-5} 
 & 0.6& 0.4& 72.3 & 31\\
\cline{2-5} 
 & 0.6& 0.6& 71.3 & 29\\
\cline{2-5} 
& 0.6& 0.8& 70.8 & 17\\
\hline
\end{tabular}
\label{ab_lam}
\end{center}
\end{table}

\begin{figure*}[t]
\centering
        \hspace{-4.5mm}
	\subfloat[]{\includegraphics[width = 0.28\textwidth]{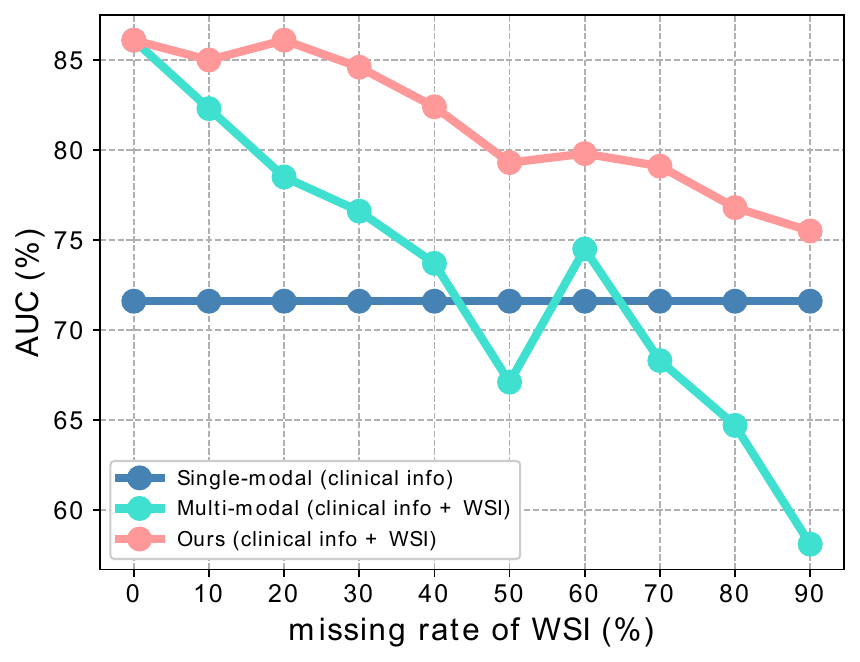}}
        \hspace{1.4mm}
	\subfloat[]{\includegraphics[width = 0.28\textwidth]{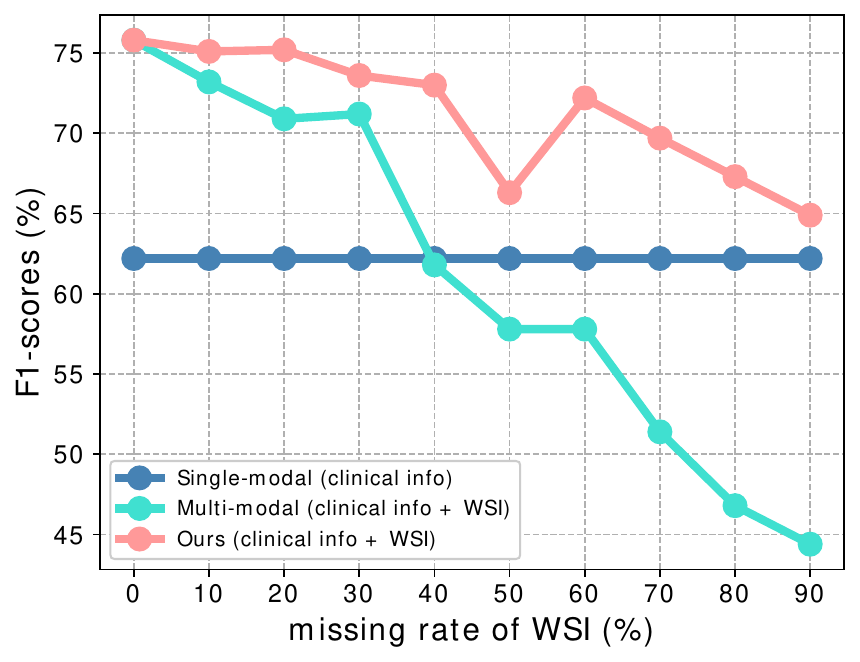}}
        \hspace{2mm}
        \subfloat[]{\includegraphics[width = 0.28\textwidth]{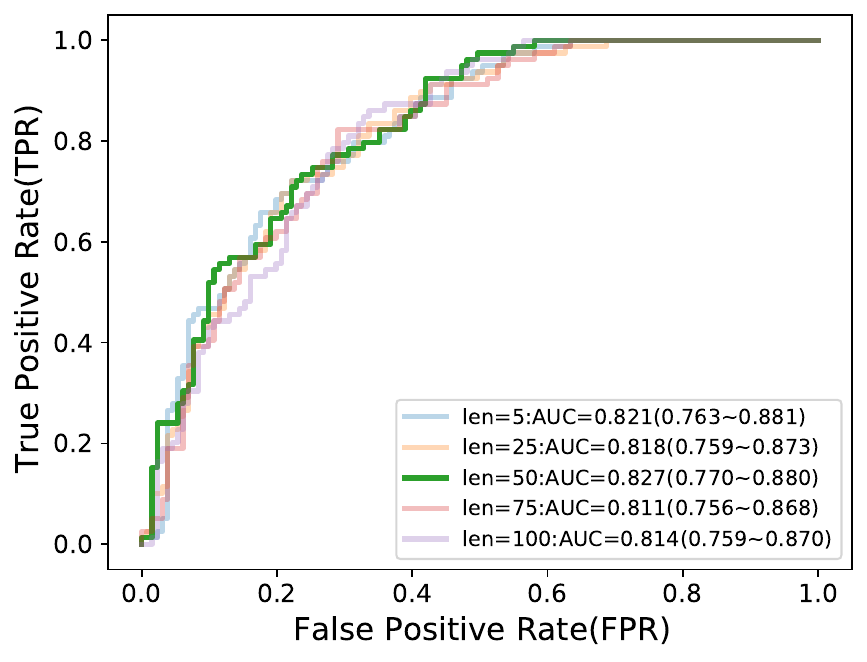}}
\caption{(a)-(b) Model performance (AUC and F1 score) at different missing rates of whole slide images. The green line is the result of the baseline method $Filling$. (c) The ROC curve, AUC value, and the confidence interval (97.5\%) under different initial lengths of the learnable prompt.}
\label{mix}
\end{figure*}

\begin{figure*}[t]
\centering
        \hspace{1mm}
	\subfloat[WSI features of single-modal model]{\includegraphics[width = 0.255\textwidth]{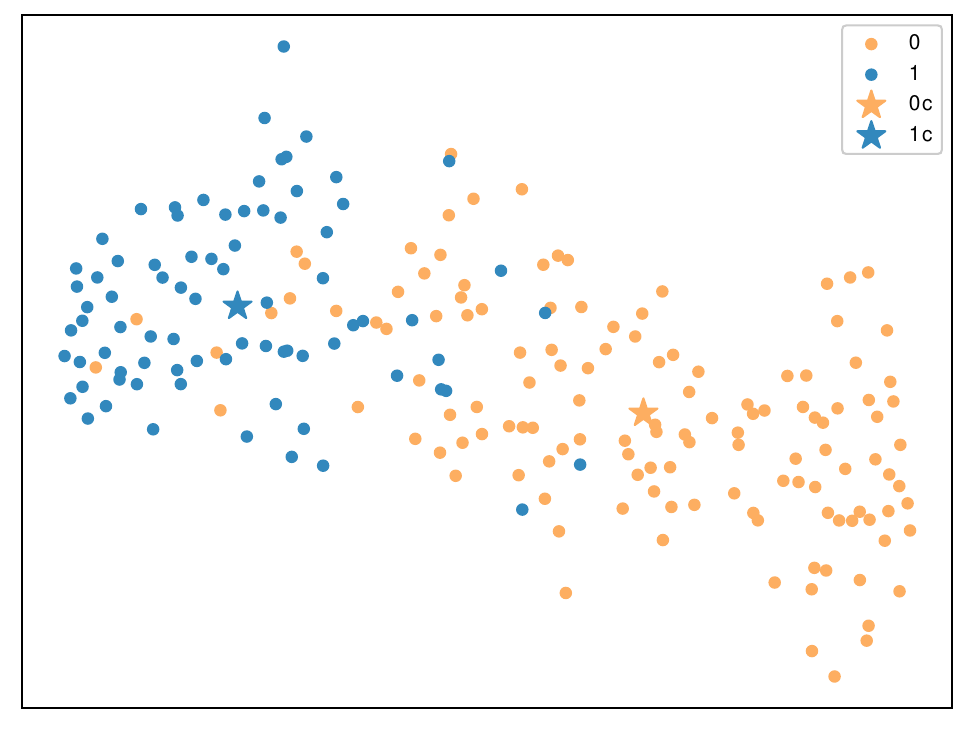}}
        \hspace{6mm}
	\subfloat[Expanding 20 times]{\includegraphics[width = 0.255\textwidth]{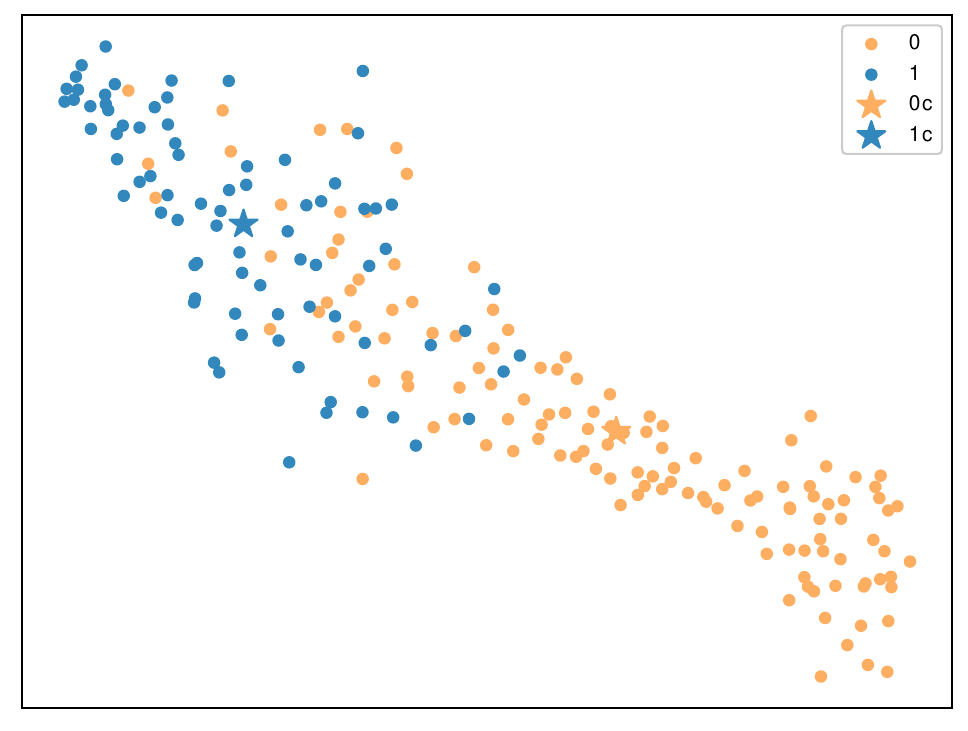}}
        \hspace{6mm}
	\subfloat[Expanding 40 times]{\includegraphics[width = 0.255\textwidth]{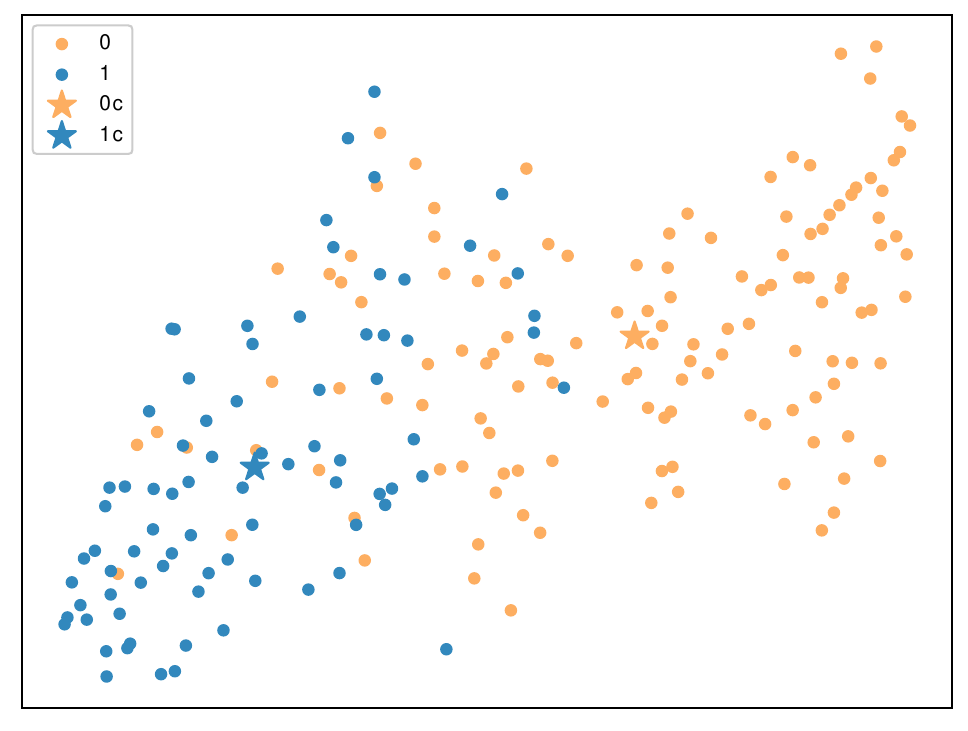}} 
\caption{Distributions of WSI features. The number `0' and `1' represent negative and positive classes, respectively. Stars mean the center of classes. WSI features from the multi-modal model are perturbed by the learning of clinical data compared to that from the single-modal model.}
\label{feas}
\end{figure*}

The results of comparisons are shown in Table \ref{comparison}. Our method achieves the best performance (bold) regarding F1 scores and AUC.
Compared to others, AE yields relatively worse performance. This method often requires a large amount of paired training data, therefore, is difficult to effectively generate the accurate features of clinical data for the prediction.
The direct filling method meets the requirement of test flexibility but does not provide valuable information about the missing clinical data. Thus, performance decreases greatly with the increase of the missing ratio and even becomes worse than that of the method with only images. The integration of two separate networks in the Ensemble method has better performance both on the complete modality and incomplete modality among the three intuitive approaches. However, the two networks are independent, and the complete modality in the training set is underutilized. For instance, a test sample only with the modality of WSI is not helped by the clinical data in the training set. Our method is also inspired by this finding and further improves the performance based on Ensemble. SMIL can also be regarded as a generative model. Differently, it is trained end-to-end, but the shared encoder may be perturbed by clinical data. 

\subsection{Ablation Study.}





\textbf{Ablation study on distillation directions}. We split the BD framework into two parts: the single-branch learning from the multi-branch ($\mathcal{M}\rightarrow \mathcal{S}$) and the multi-branch learning from the single-branch ($\mathcal{S}\rightarrow \mathcal{M}$). Then, we design the ablation study to verify the effectiveness of each part. We test in two situations with the missing ratio of 0\% (complete modality) and 80\%. F1-scores of the model with or without each part are presented in Table \ref{ablation}. We regard the independent two branches without distillation as the baseline in the ablation study. Under the test case of complete modality, the performance remains the same after adding the part $\mathcal{M}\rightarrow\mathcal{S}$  due to turning off the single branch when testing. But there is a substantial improvement after appending the part $\mathcal{S}\rightarrow\mathcal{M}$. In the case of missing modality, it is exactly the other way around. The part $\mathcal{M}\rightarrow\mathcal{S}$ is crucial to the performance of the model, while $\mathcal{S}\rightarrow\mathcal{M}$ has little effect on it. Thus, $\mathcal{M}\rightarrow\mathcal{S}$ and $\mathcal{S}\rightarrow\mathcal{M}$ are necessary for the incomplete and complete modality respectively. 

\textbf{Ablation study on the initial length of learnable prompt}. We opt for the scenario where 100\% of clinical data is missing for comparison. 
As shown in the subfigure (c) of Fig.\ref{mix}, the model performs best when the initialization length is 50 (the green line). Too long initialization of the prompt may result in memory redundant information of missing modality. We believe that shorter initializations might convey less information, yet the prompt can still serve as a reminder to the model regarding the absence of the modality. The performance will not drop significantly.

\textbf{Ablation study on $\lambda_m$ and $\lambda_s$}. Missing 100\% clinical data is considered in this experiment. We first fix $\lambda_s$ and vary $\lambda_m$ to record model performance. Then we choose the best $\lambda_m$ and change $\lambda_s$ to various values. As shown in TABLE \ref{ab_lam}, larger values of $\lambda_m$ lead to better performance, illustrating the necessity of distilling WSI features from the single-modal branch. It may overwhelm the classification loss $\mathcal{L}_{sgl}^c$ as $\lambda_s$ increases, resulting in performance degradation. From the values of the saved epoch, the model converges faster when $\lambda_s$ is larger.

\subsection{Feature Analysis between Single-modal and Multi-modal Models.}
For further study, we analyze the deep features of WSI before and after the addition of clinical information. We train a single-modal model with only WSIs and a multi-modal model with completely paired data, including WSIs and clinical data. Then, the deep features of WSI from the two models are collected. We perform feature dimensionality reduction based on t-SNE\cite{van2008visualizing} and visualize these features on the two-dimensional plane as shown in Fig. \ref{feas}. (a): The WSI features are extracted by the model trained with only WSIs. (b): The WSI features are from the intermediate output of a trained multi-modal model, in which the deep features of clinical data are expanded by 20 times compared to the original dimension. (c): The WSI features are also collected from the multi-modal model, where the deep features of clinical data are expanded by 40 times. 

We find that the WSI features from the single-modal model are more aggregated and divisible. And there is a sign that we may get worse WSI features as the feature dimension of clinical data increases. Thus, we conclude that the addition of clinical information might affect the representation learning of WSIs. Inspired by this finding, we keep the part that the multi-modal branch learns from the single-modal branch ($\mathcal{S}\rightarrow\mathcal{M}$).

\subsection{Further Investigation into the Absence of WSIs}

To validate the efficacy of our model, we consider the scenario of WSI absence. We employ the learnable prompt to alert the model of the presence of missing modality and memorize the information of WSIs from the multi-modal branch. The image encoder is removed from the single-modal branch, and the prompt is directly mapped to the deep feature. The subfigure (a) and (b) of Fig.\ref{mix} illustrate that our model far outperforms the base model ($Filling$), and it consistently outperforms the single-modal model at various missing rates. This demonstrates the effectiveness of our model no matter which modality is missing.

\section{Conclusion}

Combining modalities can improve the performance of deep learning models in the diagnosis of axillary lymph node metastasis.
However, there usually exists missing modality during test. In this paper, we propose a bidirectional distillation framework to cope with the problem of missing clinical data flexibly. Our model makes full use of the complete modality in the training set effectively via the interaction of the two branches (single-modal and multi-modal branches). The experiment results show that our model makes significant improvements at different missing rates of clinical information. Our method is model- and task-agnostic. We will further explore the effectiveness of our model in other multi-modal tasks in the future.


\bibliographystyle{IEEEtran}
\bibliography{N}

\end{document}